\documentclass[aps,prb,twocolumn,amssymb,floatfix,superscriptaddress,showpacs]{revtex4-1}

\usepackage{graphicx,color}
\usepackage{dcolumn}
\usepackage{amsmath}
\usepackage{amssymb}

\usepackage[dvipdfm]{hyperref}
\hypersetup{colorlinks=true,linkcolor=blue,urlcolor = blue, pagebackref=true,
  implicit=true,breaklinks=true,pagebackref=true,backref=true,
  bookmarks=true,bookmarksnumbered=true,hyperfootnotes=true,debug=true,
  naturalnames=false,citecolor=blue,pdfview=FitH,pdfstartview=FitH,hyperindex=true}

\begin{document}

\preprint{}

\title{NMR Evidence of anisotropic Kondo liquid behavior in CeIrIn$_5$}

\author{A.C. Shockley}
\email{acshoc@gmail.com}
\altaffiliation{Previously Department of Physics, University of California, Davis, CA 95616}
\affiliation{Laboratoire de Physique des Solides, Universit\'{e} Paris-Sud 11, UMR CNRS 8502, 91405 Orsay, France}

\author{K.R. Shirer}
\author{J. Crocker}
\author{A.P. Dioguardi}
\author{C.H. Lin}
\author{D.M. Nisson}
\author{N. apRoberts-Warren}
\author{P. Klavins}
\author{N.J. Curro}
\affiliation{Department of Physics, University of California, Davis, CA 95616}


\date{\today}

\begin{abstract}
We report detailed Knight shift measurements of the two indium sites in the heavy fermion compound CeIrIn$_5$ as a function of temperature and field orientation.  We find that the Knight shift anomaly is orientation-dependent, with a crossover temperature $T^*$ that varies by 50\% as the field is rotated from (001) to (100).  This result suggests that the hybridization between the Ce 4f states and the itinerant conduction electrons is anisotropic, a result that reflects its collective origin, and may lead to anisotropic Kondo liquid behavior and unconventional superconductivity.
\end{abstract}


\maketitle


\section{Introduction}

Heavy electron materials exhibit a number of interesting correlated electron phenomena, including unusual broken symmetry ground states, quantum criticality and non-Fermi liquid behavior, which arise from the interactions between  a lattice of nearly-localized 4f electrons and itinerant conduction electron states.\cite{doniach,hewson1997kondo}  When the 4f states are weakly hybridized with the itinerant states, the materials tend to exhibit long-range antiferromagnetism mediated by RKKY interactions;  in the opposite limit the long-range order disappears, the resulting itinerant quasiparticles have enhanced effective masses, and the system typically is unstable towards unconventional superconductivity.\cite{zachreview,ColemanHFdeath,ColemanQCreview}  {The emergence of a heavy-fermion fluid in close proximity to an antiferromagnetic instability of localized moments remains an active area of experimental and theoretical research.} Several Ce-based compounds happen to exhibit a level of hybridization that places them close to the quantum critical (QC) boundary between long-range antiferromagnetism and superconductivity. As a result small perturbations induced by doping or pressure can result in dramatic changes to the ground state properties.\cite{tuson,ParkDropletsNature2013}  These compounds offer an ideal testing ground to investigate the interplay between the hybridization and the emergent states of the strongly correlated system.

\begin{figure}
\centering
\includegraphics[width=\linewidth]{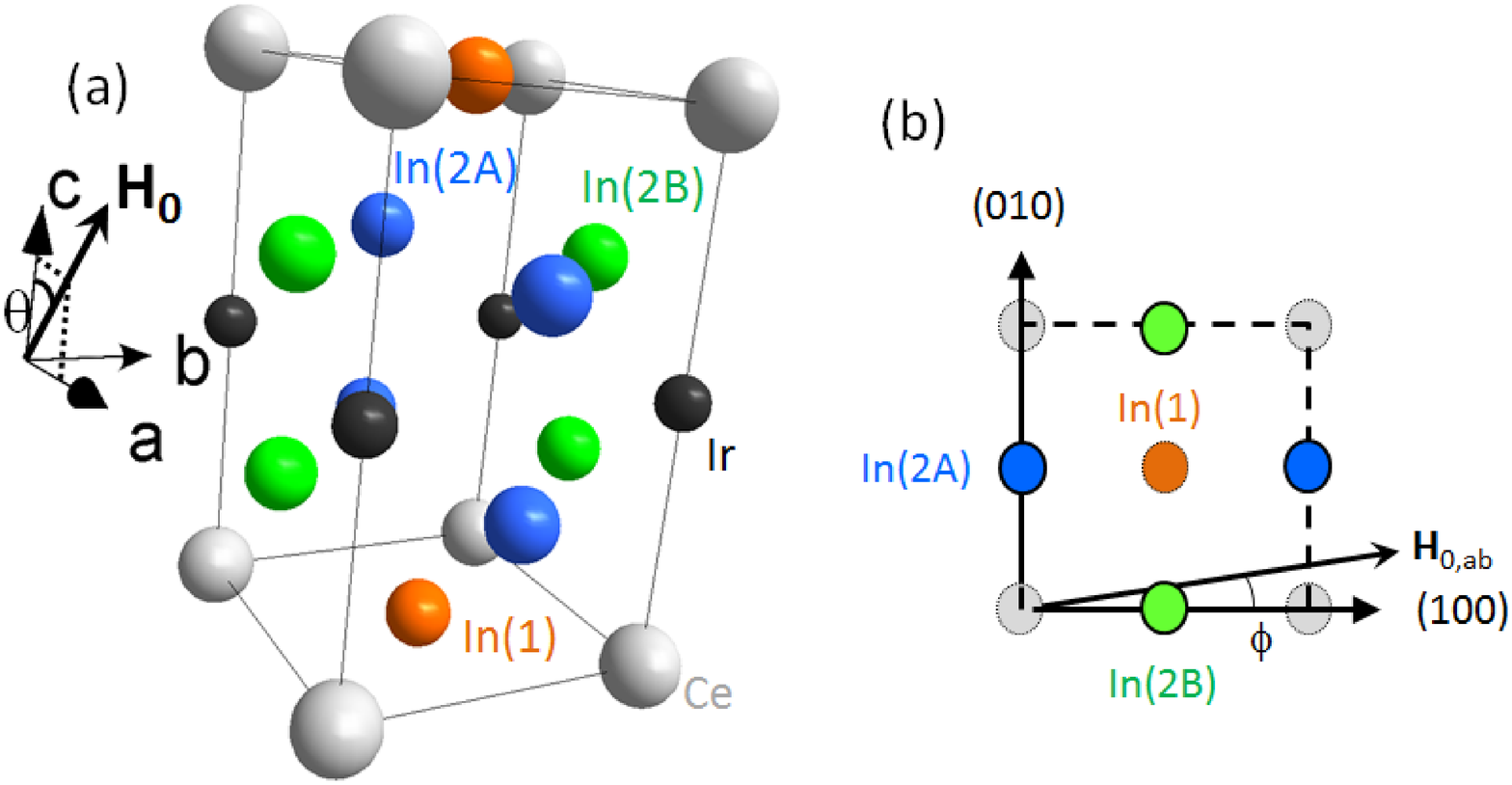}
\includegraphics[width=\linewidth]{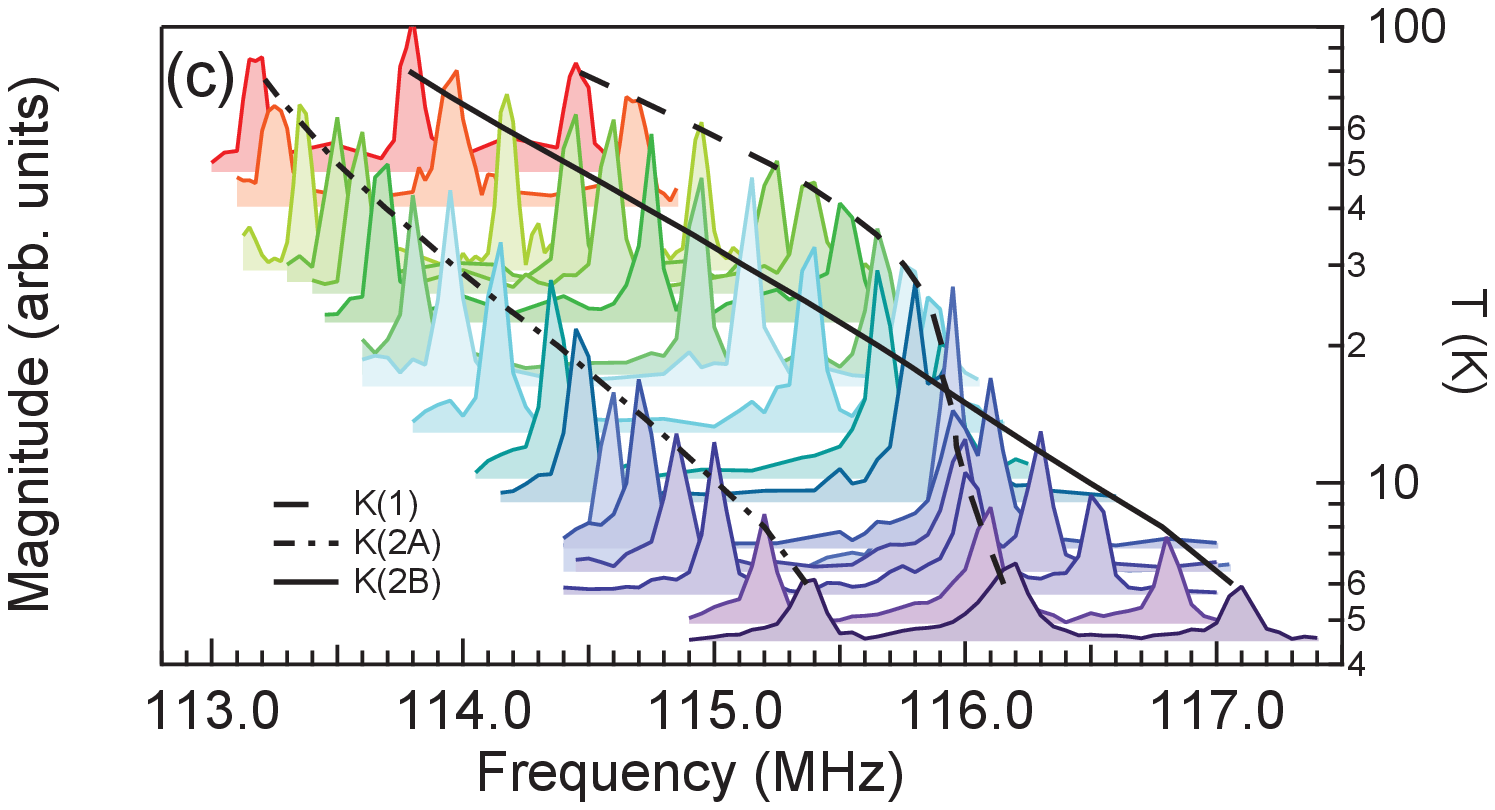}
\caption{(a) Unit cell indicating the three indium sites and the field orientation. (b) Projection of the unit cell in the $ab$ plane. (c)
A representative frequency swept spectrum of CeIrIn$_5$ at 11.7T for $\theta=44^{\circ}$ and temperatures from 6K to 80K; the magnitude is normalized by temperature. The temperature-dependent Knight shifts for the In(1), In(2A) and In(2B) sites are scaled over the raw spectral data. \label{spectra}}
\end{figure}

CeIrIn$_5$ is an excellent example of a system close to a QC boundary; while it is superconducting below 0.4K, the normal state exhibits antiferromagnetic fluctuations and non-Fermi liquid behavior.\cite{ceirindiscovery, ZhengIr115PRL} {Thus this compound can provide vital information about the emergence of the coherent heavy-fermion fluid near a QC boundary.} Dynamical mean field theory (DMFT) calculations  indicate that CeIrIn$_5$ undergoes a crossover from localized to itinerant electron behavior with decreasing temperature, accompanied by changes to the Fermi surface.\cite{HauleCeIrIn5,Choi2011}  Experimental evidence is provided by resistivity, specific heat, and nuclear magnetic resonance (NMR) Knight shift measurements which are well-described by a two-fluid picture of heavy fermion behavior.\cite{YangPinesNature}
Recent calculations have shown that this hybridization-driven crossover is strongly anisotropic in this material.\cite{haule2010dynamical}  {Here we  provide direct experimental evidence for such hybridization anisotropy, which may play a key role in stabilizing the unconventional superconductivity in this family of heavy fermions.}
{The DMFT calculations indicate that}  since the local 4f states are anisotropic,  the hybridization is dominated by the orbital overlap between the Ce 4f and the out-of-plane In(2) electron orbitals (see Fig. \ref{spectra}). This hybridization should be manifest in the spin susceptibility $\chi_{c\!f}$, describing the correlations between the itinerant and local moment electron spins. This quantity  can be directly probed via Knight shift experiments.\cite{Curro2004,ShirerPNAS2012}  We have conducted detailed angular-dependent studies of the In(1) and In(2) Knight shifts, and find that $\chi_{c\!f}$ indeed depends on the orientation of the applied magnetic field with respect to the crystal axes.  The temperature dependence of this correlation  function is determined by the Kondo lattice coherence temperature, $T^*$, which we find to be largest along the Ce-In(2) bond axis.

\section{Knight Shift Measurements}

High quality single crystals of CeIrIn$_5$ were synthesized using the standard flux method described in Ref. \onlinecite{Moshopoulou200125}. Characterization with powder X-ray diffraction showed the samples were pure with a small amount of In flux.\cite{Shockley2011} A large single crystal with dimensions 3mm $\times$ 3mm $\times$ 1mm was chosen for the NMR studies. NMR measurements were performed in an Oxford high-homogeneity NMR magnet at a fixed field of 11.7T. All spectra were obtained using a standard Hahn echo pulse sequence.\cite{slichter1990principles} The orientation of the sample was controlled by a single-axis goniometer, and the sample was mounted such that the applied field was directed at an angle $\theta$ from $(001)$, in the plane spanned by (100) and (001), as shown in Fig. \ref{spectra}(a).  For each angle, a full spectrum including several different satellite  transitions of the $^{115}$In ($I=9/2$) was obtained using an automated tuning system integrated with the NMR spectrometer.  The quadrupolar nature of this isotope enabled us to extract the orientation of the field, and hence the Knight shift, as described in detail in the Appendix.  There are four In(2) sites per unit cell, and when $\theta>0$ these four sites split into two inequivalent sites depending on whether the field is parallel or perpendicular to the face of the unit cell (see Fig. \ref{spectra}(b)). We refer to these two In(2) sites as In(2A) and In(2B). Characteristic spectra of the In(1), In(2A) and In(2B) sites are shown in Fig. \ref{spectra}(c) for $\theta = 44^{\circ}$ at several different temperatures.   Each of the three sites clearly exhibits  different temperature dependent behavior.  Detailed spectra were also measured at various rotation angles in order to observe any anisotropy in the temperature dependence.

\begin{figure}
\centering
\includegraphics[width=\linewidth]{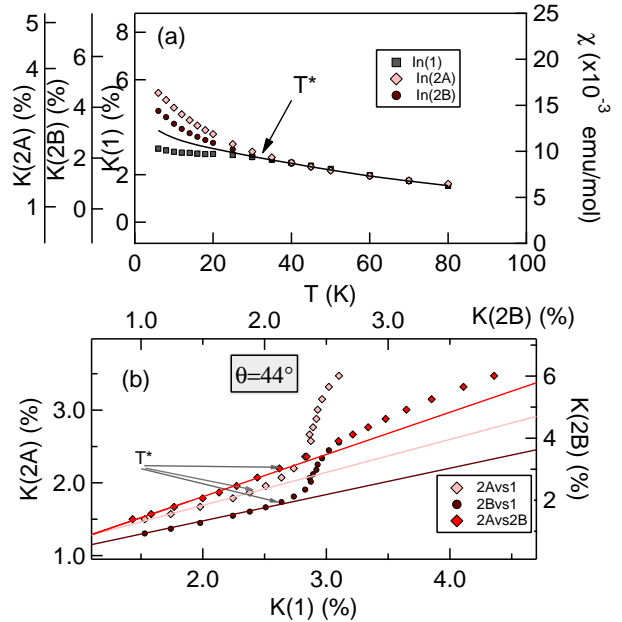}
\caption{(a) Knight shift  at $\theta = 44^{\circ}$ (solid points) and bulk susceptibility $\chi(T) = \chi_c\cos^2(\theta) + \chi_a\sin^2(\theta)$ versus temperature (solid line). (b) $K(2A)$ and $K(2B)$ versus $K(1)$, and $K(2A)$ versus $K(2B)$ with temperature implicit. Solid lines are fits to the high temperature ($T>T^*$).  $T^*$ is the temperature below which the linear relationship between these quantities breaks down. \label{KvsK}}
\end{figure}

For a spin 1/2 nucleus, the resonance frequency is given by $\omega = \gamma H_0 (1+ K(\theta,\phi))$, where $\gamma$ is the gyromagnetic ratio, $H_0$ is the magnetic field, and $K(\theta,\phi) = \mathbf{H}_0\cdot \mathbb{K}\cdot\mathbf{H_0}/H_0^2$. Here $\mathbb{K}$ is the Knight shift tensor, with principal axes lying along the unit cell directions. In general, the Knight shift arises because of the hyperfine coupling between the nuclear and electron spins of the material, which gives rise to an effective hyperfine field at the nuclear site in addition to the external field, thus shifting the resonance frequency.  Hyperfine couplings can arise from on-site Fermi contact interactions, as well as via transferred couplings to electron spins located on neighboring atoms.   The exact values of these couplings depend on details of the electronic structure of the material, are different for each site, and are generally difficult to compute.  However, it is useful to consider an effective hyperfine interaction that is appropriate for heavy fermion materials: $\mathcal{H}_{hyp} = \mathbf{\hat{I}}\cdot (\mathbb{A}\cdot \mathbf{S}_c  + \mathbb{B}\cdot \mathbf{S}_f)$, where $\mathbb{A}$  and $\mathbb{B}$ are temperature-independent  hyperfine coupling tensors to the conduction electron and local moment spins, $\mathbf{S}_c$ and $\mathbf{S}_f$.\cite{curro2004scaling} $\mathbf{\hat{I}}$ is the nuclear spin on the ligand site, in this case either the In(1), the In(2A), or the In(2B). In the paramagnetic state, the spins are polarized by the external field, and the Knight shift is given by $\mathbb{K} = \mathbb{K}_0 + \mathbb{A}\cdot{{\chi}}_{cc} + (\mathbb{A}+\mathbb{B})\cdot\chi_{c\!f} + \mathbb{B}\cdot\chi_{f\!f}$, where $\chi_{ij} = \langle S_{i}S_{j}\rangle$ are the  components of the total susceptibility $\chi = \chi_{cc} + 2\chi_{cf} + \chi_{f\!f}$, and $\mathbb{K}_0$ is the temperature-independent orbital shift tensor.  For sufficiently large temperatures $\chi_{f\!f}$ is the dominant contribution thus $\mathbb{K} \approx \mathbb{K}_0 + \mathbb{B}\cdot\chi$. As a result, $K_{\alpha}$ is linearly proportional to $\chi_{\alpha}$, where  $\alpha  = (a,b,c)$ are the principal directions of the tensor.  Furthermore, since the shift of each site is proportional to $\chi$, each shift is also proportional to the shifts of the other sites, as shown in detail in the Appendix. This linear dependence is evident in Figs. \ref{KvsK} for $T > T^*$.

\begin{figure}[h]
\includegraphics[width=\linewidth]{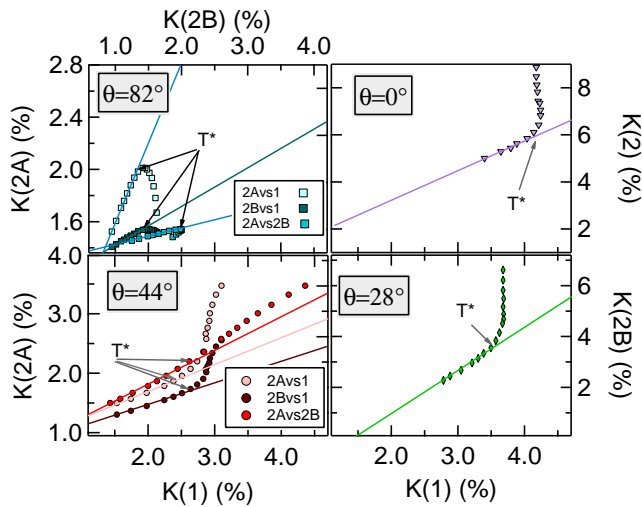}
\caption{Clockwise from the top right pane: $K(2)$ ($K(2A) = K(2B) = K(2)$ for this orientation) vs. $K(1)$ for $\theta=0^{\circ}$ (purple inverse triangles), $K(2B)$ vs. $K(1)$ for $\theta=28^{\circ}$ (green diamonds), $K(2A)$ and $K(2B)$ vs. $K(1)$ and $K(2A)$ vs. $K(2B)$ for $\theta=44^{\circ}$ (pink and dark red and red circles, respectively), and $K(2A)$ and $K(2B)$ vs. $K(1)$ and  $K(2A)$ vs. $K(2B)$ for $\theta=82^{\circ}$ (light blue, dark blue and blue squares, respectively). Solid lines are fits to the high temperature portion as described in the text. $T^*$ is indicated by the grey arrows. \label{KvsKv2}}
\end{figure}

Below the coherence temperature, $T^*$, the conduction and local moment spin degrees of freedom become entangled, and $\chi_{c\!f}$ grows in magnitude relative to $\chi_{\!f\!f}$.  As a result,  $K_{\alpha}$ is no longer proportional to $\chi_{\alpha}$, as seen in Fig. \ref{KvsK}.   $T^*$ is a material-dependent crossover temperature that depends on the hybridization and intersite couplings between the $S_f$ spins in the Kondo lattice.\cite{YangPinesNature,YangPinesPNAS2012,HauleCeIrIn5,jiang2014universal} $T^*$ can be measured experimentally via independent measurements of $K_{\alpha}$ and $\chi_{\alpha}$: when $K_{\alpha}$ is plotted versus $\chi_{\alpha}$ with temperature as an implicit variable, the linear relationship breaks down at $T^*$, as observed in Fig. \ref{KvsK} at $\theta = 44^{\circ}$.  Several other pairs of shifts and angles are shown in Fig. \ref{KvsKv2}, and in each case there is a clear break in this linear relationship at low temperatures.

\begin{figure}
\centering
\includegraphics[width=0.98\linewidth]{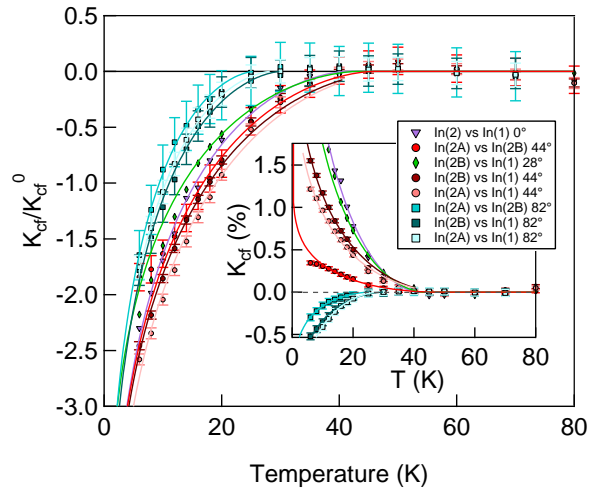}
\caption{$K_{cf}/K_{cf}^0$ versus $T$ for various angles  (colors and symbols are defined as in Fig. \ref{KvsK}). The solid lines are fits as described in the text. INSET: $K_{cf}$ versus $T$.  \label{Kcf}}
\end{figure}

In order to discern the influence of hybridization anisotropy, it is important to measure $T^*$ as a function of angle. Our previous studies of the In(1) site in CeIrIn$_5$ indicated that $T^* \sim 40$ K, and did not appear to vary significantly with field orientation or magnitude.\cite{ShirerPNAS2012,CeIrIn5HighFieldNMR}  However, the precision of the $T^*$ measurement is limited for the In(1) site because the  coupling constants $A_a = B_a$  in the plane. Therefore the magnitude of the Knight shift anomaly gradually decreases with angle and vanishes for $\mathbf{H}_0 \parallel (100)$.  This problem can be circumvented by measuring the Knight shifts of both of the In(2) sites and the In(1) site.  This approach is superior because all of the Knight shift measurements can be acquired simultaneously at the same crystalline orientation without the need for separate measurements of $\chi$.\cite{CeIrIn5HighFieldNMR} The behavior below $T^*$ is governed by the temperature dependence of the correlation function $\chi_{c\!f}(T)$.  As the the conduction electron and local moments become entangled, this quantity grows in magnitude and can be extracted from the Knight shift below $T^*$. To do so, we fit the high temperature data ($T > T^*$) for each pair $(K_1, K_2)$ of Knight shifts to $K_1 = a + b K_2$, and then plot $K_{c\!f}(\theta, T) = K_1(T,\theta) - a-b K_2(T,\theta)$ versus temperature in the inset of Fig. \ref{Kcf}. As shown in the Appendix, this quantity is proportional to $\chi_{c\!f}(\theta,T)$ and becomes non-zero below $T^*$.   The constants $a$ and $b$ depend on the ratios of hyperfine couplings of the various pairs of sites and are unimportant for our analysis.\cite{CeIrIn5HighFieldNMR}  We have confirmed that these constants are consistent for three different data sets.

\section{Anisotropy}

As seen in Fig. \ref{Kcf}, $K_{c\!f}$ vanishes above $T^*$, but  grows in magnitude with decreasing temperature below this temperature. This data clearly indicate  that the onset temperature, $T^*$, depends on the angle $\theta$. {This angular variation is model independent, and can be discerned both in the plots of $K_i$ versus $K_j$ in Figs. \ref{KvsK} and \ref{KvsKv2}.  For concreteness }we  fit the temperature dependence of $K_{c\!f}$ to the two-fluid expression\cite{YangDavidPRL}:
\begin{equation}
K_{c\!f}(T) = K_{c\!f}^0(1-T/T^*)^{3/2}[1+\ln(T^*/T)],
\label{eqn:Kcf}
\end{equation}
and plot $K_{c\!f}^0$ and $T^*$ versus angle $\theta$ in Figs. \ref{TstarVSangle}(a) and \ref{TstarVSangle}(b). $K_{c\!f}^0$ is proportional to a complex ratio of the hyperfine couplings {and anisotropic g-factors of the material}, and the angular dependence of this quantity seen in Fig. \ref{TstarVSangle}(a) reflects the anisotropies of these couplings. {The main panel of Fig. \ref{Kcf} shows $K_{cf}(T)$ normalized by $K_{c\!f}^0$, which removes any anisotropies introduced by the hyperfine couplings and g-factors. The onset temperature of the anomaly, $T^*$, varies with angle.  Here $T^*$ is unrelated to the hyperfine couplings and reflects intrinsic properties of the electronic degrees of freedom of the Kondo lattice. As seen in Figs. \ref{Kcf} and \ref{TstarVSangle}(b), as the field angle rotates from the $(001)$ direction, $T^*$ increases from 40K to nearly 50K at $44^{\circ}$ and then reaches a minimum of 26 K for the  $(100)$ direction.   In order to parameterise this anisotropy, we fit this angular dependence to the form $T^*(\theta) = T^*_0 + T^*_2\cos(2\theta) + T^*_4\cos(4\theta)$, which qualitatively reproduces the hybridization function calculated in Ref. \onlinecite{haule2010dynamical}. We find $T^*_0 = 42(2)$ K, $T^*_2= -7(2)$ K and $T^*_4 = 7(2)$ K, shown in Fig. \ref{TstarVSangle}(c). These results reveal that the heavy electron fluid, which emerges from the collective hybridization of the lattice of 4f sites with the conduction electrons, is anisotropic in this material. This result suggests that the hybridization is not isotropic and has four-fold symmetry.

\begin{figure}
\centering
\includegraphics[width=\linewidth]{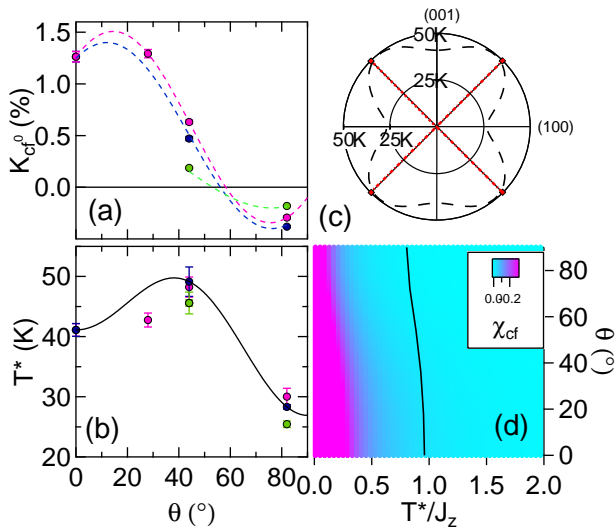}
\caption{(Color online) (a) $K_{cf}^0$ and (b) $T^*$ versus angle as determined from the fits shown in Fig. \ref{Kcf}, as determined by plotting $K(2B)$ vs. $K(1)$ (pink), $K(2A)$ vs. $K(1)$ (blue) and $K(2A)$ vs. $K(2B)$ (green).  Dashed lines are guides to the eye, and the solid line in (b) is a fit as described in the text. (c)  $T^*(\theta)$  shown as a polar plot, relative to the $(001)$ (vertical) and $(001)$ (horizontal) directions.  The dotted red lines indicate the Ce-In(2) directions. (d) $\chi_{c\!f}(T,\theta)$ and $T^*/J_z$ (solid line) versus $\theta$ for the two-spin model discussed in the text. \label{TstarVSangle}}
\end{figure}

\subsection{Hybridization}

A recent  analysis of data in a broad range of heavy fermion materials indicated that $T^*$ is proportional to the intersite RKKY exchange interaction, which itself is proportional to $J^2$, where the Kondo coupling $J$ is a function of the hybridization.\cite{YangPinesNature} Therefore, anisotropy in the hybridization should be reflected in the experimentally measured quantity, $T^*$. In order to discern how an anisotropic hybridization can give rise to anisotropy  in the susceptibility $\chi_{cf}$, it is instructive to consider a generalization of the two-spin model introduced in Ref. \onlinecite{ShirerPNAS2012}.   We consider an anisotropic coupling between two free spins, $\mathbf{S}_c$ and $\mathbf{S}_f$: $\mathcal{H} = J_{\perp}({S}_c^x{S}_f^x+{S}_c^y{S}_f^y)+J_z{S}_c^z{S}_f^z$, where $J_{z,\perp}$ is the coupling between the spins derived from the anisotropic hybridization parallel (perpendicular) to the z-axis. This model is the single-site limit of the periodic Anderson model in the limit of large on-site repulsion, $U$, relative to the hybridization, $V$, such that $J_\alpha  = 4V_{\alpha}^2/U$.\cite{jiang2014universal}  In this case, the susceptibilities $\chi_{cc}$, $\chi_{cf}$ and $\chi_{ff}$ are exactly solvable.  For the isotropic case $J_{\perp}=J_z$, the $\chi_{ij}$ are all isotropic and scale as $T/T^*$, where $T^* = J_z/k_B$.   When $J_{\perp}\neq J_z$, these susceptibilities become anisotropic tensors, such that the susceptibility becomes angular dependent: $\chi_{cf}(T,\theta) = \chi_{cf}^z(T)\cos^2(\theta) + \chi_{cf}^{\perp}(T)\sin^2(\theta)$,  shown in Fig. \ref{TstarVSangle}(d) for the case  $J_{\perp}=0.2 J_z$.  We fit this quantity to  Eq. \ref{eqn:Kcf} for several values of $\theta$  and the solid line in Fig. \ref{TstarVSangle}(d) shows the  fitted values of $T^*(\theta)$.  Clearly $T^*$ is anisotropic, although this model not sophisticated enough to capture the four-fold variation observed in Fig. \ref{TstarVSangle}(b). A model including multiple sites would represent the full lattice better and be more likely to resemble the experimental measures.

\subsection{Crystalline Electric Field}


\begin{figure}[h]
\includegraphics[width=\linewidth]{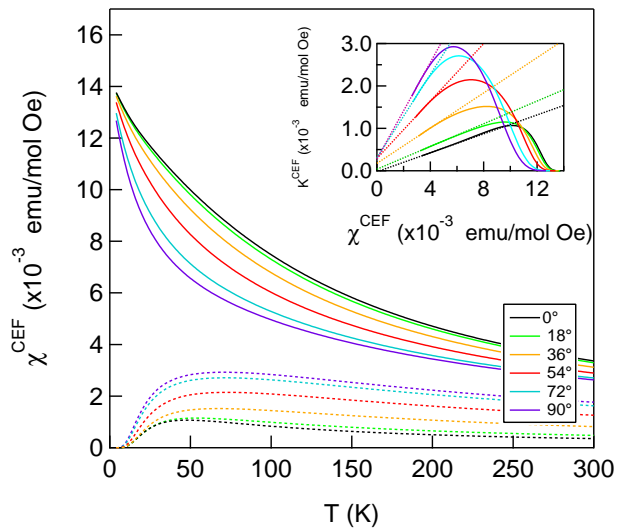}
\caption{Susceptibility ($\chi^{CEF}$, solid lines) and Knight shifts ($K^{CEF}$, dotted lines) versus temperature calculated in the CEF model for various field orientations, as described in the text.  INSET:  $K^{CEF}$ versus $\chi^{CEF}$ (solid lines) for the same orientations.  Dashed lines are fits to the high temperature data points.  \label{fig:CEFKandChi}}
\end{figure}

An alternative interpretation of Knight shift anomalies is that the hyperfine coupling constants depend on the particular crystalline electric field (CEF) doublets.\cite{KnightShiftAnomalyCEF,Curro2001} The strong spin-orbit coupling combined with CEF interactions at the Ce ions give rise to a temperature-dependent anisotropic $g$-factor. The Ce$^{3+}$ ions in this material experience a CEF that splits the $J=5/2$ ground state multiplet into three doublets, with excited states energies  $\Delta_1 = 6.7$ meV and $\Delta_2=29$ meV above the ground state \cite{CEF115study,WillersCEF115s}.  In order to explore the possible role of the CEF in the anisotropy we observe in $K_{cf}$, we have computed $K_{cf}^{CEF}$ and $\chi^{CEF}$ as a function of field orientation using the hyperfine coupling model discussed in Refs. \onlinecite{CeIrIn5HighFieldNMR,KnightShiftAnomalyCEF,Curro2001}.  In this scenario, the hyperfine coupling between the In(1) site and the Ce spin depends on the particular CEF doublet; thus when the temperature $T \lesssim \Delta_1/3 k_B$, the thermal population of the excited states is significantly reduced and the effective hyperfine field changes.  As a result, the Knight shift differs from the susceptibility below the  anomaly temperature $T^*_{CEF} \sim \Delta_1/k_B$.  Here we computed $\chi_{c,ab}^{CEF}$ and  $K_{c,ab}^{CEF}$ using the same CEF parameters and hyperfine couplings as in Ref. \onlinecite{CeIrIn5HighFieldNMR} in a field of 11.7 T. These quantities are shown in  Fig. \ref{fig:CEFKandChi}.  Note that these calculations do not accurately capture the behavior of the real material because this model neglects the role of hybridization of the Ce 4f states.  Nevertheless, there is a clear anisotropy in the magnitudes of both $K^{CEF}$ and $\chi^{CEF}$, which reflects the anisotropy of the $g$-factor of the Ce.  We have also assumed isotropic hyperfine couplings in this calculation, but relaxing this assumption would simply modify the relative scale factors of the Knight shifts shown in Fig. \ref{fig:CEFKandChi}.

\begin{figure}[h]
\includegraphics[width=\linewidth]{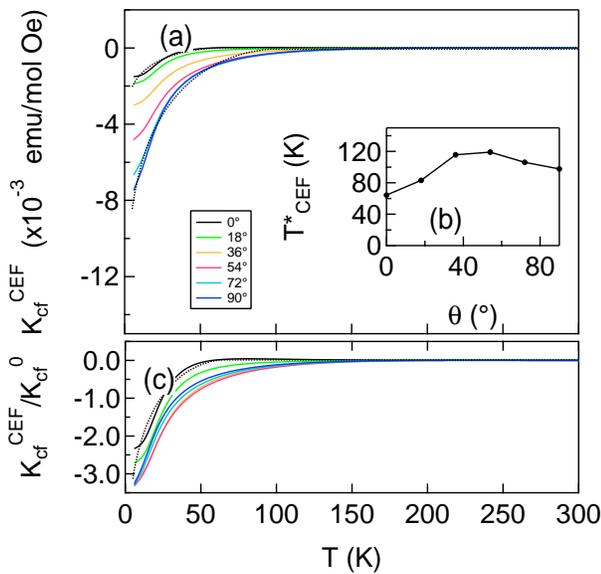}
\caption{(a) $K_{cf}^{CEF}$ versus $T$ for several different field orientations (see legend in Fig. \ref{fig:CEFKandChi}) using the fits to the high-temperature $K^{CEF}$ versus $\chi^{CEF}$ relationship.  (b) $T^*_{CEF}$ extracted by fitting the theoretical $K_{cf}^{CEF}$ results to Eq. 1 in the main text as a function of angle. (c) $K_{cf}^{CEF}/K_{cf}^0$ versus $T$  for various angles.  The dotted line is a fit to Eq. 1 in the main text. \label{fig:CEFKcf}}
\end{figure}

We then fit the high temperature portion ($T>100$ K) of $K^{CEF}(\theta)$ versus $\chi^{CEF}(\theta)$ for various values of $\theta$, and plotted $K_{cf}^{CEF}(\theta) = K^{CEF}(\theta) - a - b\chi^{CEF}(\theta)$, shown in Fig. \ref{fig:CEFKcf}(a).  $\chi^{CEF}(\theta)$ grows in magnitude below $T_{CEF}^* \approx 55 K \sim 0.7 \Delta_1$, and varies strongly with $\theta$.   For concreteness, we fit
$K_{cf}^{CEF}(\theta)$ to  Eq. \ref{eqn:Kcf} to extract $T^*_{CEF}$,  shown in Fig. \ref{fig:CEFKcf}(b).  However, the fits, shown for $\theta = 0$ and $90^{\circ}$ as dotted lines in  Fig. \ref{fig:CEFKcf}(a), are poor and do not capture well the detailed temperature dependence of the theoretical CEF curves. The fitted $T^*$ increases from 64 K to 120 K and then down to 98 K as the field varies from the $c$-axis to the $ab$-plane. Fig. \ref{fig:CEFKcf}(c) shows $K_{cf}^{CEF}(\theta)$  normalized by $K_{cf}^0$ in order to remove the anisotropy of the $g$-factor.
Although the angular dependence of $T^*$ extracted from these fits qualitatively reproduces the experimental observations, the temperature dependence of $K_{cf}^{CEF}$ does not match well with the experimental data shown in Fig. \ref{Kcf}.  From this data alone, it is not possible to rule out this model, although previous work indicates that the field dependence of the Knight shift anomaly is not captured by the CEF model.\cite{CeIrIn5HighFieldNMR}  It is possible that both CEF and hybridization effects could be playing a role in determining the anisotropic behavior we observe.
%

\section{Discussion}

An anisotropic coherence temperature has important implications in the context of the two-fluid model, and can explain various observations.  For example, in compounds such as CeCoIn$_5$ and CeCu$_2$Si$_2$, evidence suggests that the temperature onset of the Knight shift anomaly differed when the field was parallel or perpendicular to the tetragonal $c-$axis.\cite{curro2004scaling} Our experiments on CeIrIn$_5$ suggest that hybridization anisotropy may also be present in these materials, and call for further measurements. Anisotropy may also explain the variations in $T^*$ measured  by different experimental techniques such as NMR, resistivity, specific heat or Hall measurements.\cite{YangPinesNature}  If there is a magnetic field present to break the symmetry, then experiments that couple to the susceptibility will reveal this anisotropy.   On the other hand, for measurements in the absence of a field, such as resistivity or specific heat, the heavy electron fluid will have an isotropic response with a slightly different temperature scaling.  For example, in the isotropic case of the two-spin model discussed above, both the susceptibility and the specific heat scale as $T/T^*$, where $T^* = J_z/k_B =J_{\perp}/k_B$.   However, when $J_{\perp}\neq J_z$ the susceptibility scales as $T/T^*(\theta)$ and the specific heat scales as $T/T_C^*$, where $T_C^*$ is isotropic. In this case, both $T^*(\theta)$ and $T_C^*$ are more complicated functions of $J_z$ and $J_{\perp}$.

The anisotropy we observe in the static susceptibility will also be reflected in the dynamic susceptibility of the Kondo liquid, which may play a role in the emergence of superconductivity in this material.  Anisotropic spin fluctuations have been shown to give rise to d-wave superconductivity and enhance $T_c$ in for 2D fluctuations.\cite{MonthouxPinesReview}  Superconductivity appears to be fairly common in certain heavy fermion families, such as the CeMIn$_5$ series, but not in other Ce-based heavy fermion families. Our observations suggest the reason for the  stability of superconductivity in the CeMIn$_5$ series may arise from the particular orbital overlap between the In(2) and the Ce sites in this structure, giving rise to the anisotropy in $T^*$ we observe experimentally.

In summary, we have found evidence that the coherence temperature $T^*$ as measured by the Knight shift is anisotropic in CeIrIn$_5$, reflecting an anisotropic collective hybridization in the Kondo lattice among multiple sites.  Our results demonstrate that the NMR Knight shift is a vital new tool to explore and quantify this anisotropy, and suggests that the In(2) sites in this compound play a key role in the development of the heavy electron fluid. Detailed calculations,  for example Quantum Monte Carlo simulations, should be carried out in order to test the effects of anisotropic hybridization and discern whether the four-fold symmetry we observe arises from collective hybridization among multiple sites.

\section{Acknowledgements}

We thank  Y-F. Yang, D. Pines, M. Jiang and R. Scalettar for enlightening discussions. A.C. Shockley would also like to thank F. Bert, I. Mukhamedshin and P. Mendels.  Work at UC Davis was supported by the NSF under Grant No.\ DMR-1005393.

\section{Appendix}

\subsection{Spectrum Analysis}

Each spectrum, covering up to a range of 40 MHz, contains nine In(1) transitions plus up to eighteen In(2) transitions, depending on the orientation of the field with respect to the crystal.  The resonance frequencies are determined by the  NMR Hamiltonian: $\mathcal{H}_n = \gamma \hbar \mathbf{\hat{I}}\cdot\left(\mathbb{I}+\mathbb{K}\right)\cdot\mathbf{H}_0 +\mathcal{H}_Q$, where $\mathbb{K}$ is the Knight shift tensor, $\gamma$ is the gyromagnetic ratio, $\mathbf{H}_0$ is the external applied field, and $\mathcal{H}_Q$ is the quadrupolar Hamiltonian.  The latter is given by:
\begin{equation}
\mathcal{H}_Q = \frac{\hbar}{6}\left[\omega_{zz}(3\hat{I}_z^2 - \hat{I}^2) + (\omega_{xx} - \omega_{yy}) (\hat{I}_x^2 - \hat{I}_y^2)\right],
\end{equation}
where $\left(\omega_{xx},\omega_{yy},\omega_{zz}\right)$  are the eigenvalues of the electric field gradient (EFG) tensor, with eigenvectors directed along the $x,y,z$ directions.  For the In(1) site, $\omega_{zz} = 6.07$ MHz along (001), and $\omega_{xx} = \omega_{yy} = -3.04$ MHz along (100) and (010). For the In(2) $\omega_{zz} = 18.17$ MHz along (100),   $\omega_{xx} = -13.26$ MHz along (010), and $\omega_{yy} = -4.91$ MHz along (001).  $\mathcal{H}_n$ was diagonalized numerically and the resonance frequencies were fit to the spectral data with the shift $K_{\alpha\!\alpha}$, the polar angle $\theta$, and the azimuthal angle, $\phi$, left as variable parameters. The In(1) site has axial symmetry, therefore there are  nine equally-spaced satellite transitions whose frequencies only depend on $\theta$. For each orientation of the crystal, we fit the positions of the In(1) peaks in order to extract the angle $\theta$.  The azimuthal angle $\phi$ describes the orientation of $\mathbf{H}_0$ relative to (100). By analyzing the satellites of the In(2) we found $\phi = 0 \pm 2^{\circ}$ for each rotation of the goniometer.

\subsection{Relationship between shifts of different sites}

 The hyperfine interaction is given by $\mathcal{H}_{\rm hf} = \mathbf{\hat{I}}\cdot[A \mathbf{S}_c + B \mathbf{S}_f$], where $A$ and $B$ are the hyperfine couplings to the itinerant electron spins, $\mathbf{S}_c$, and to the local moment spins, $\mathbf{S}_f$.   In this case the Knight shift of each site is  given by:
\begin{equation}
K_{i} = K_i^0 + A_i\chi_{cc} + (A_i+B_i)\chi_{c\!f} + B_i\chi_{f\!f}
\label{eqn:Knightshift}
\end{equation}
where $i$ corresponds to In(1), In(2A) or In(2B), $K_i^0$ is a temperature independent orbital term, and the components of the susceptibility are given by $\chi_{\alpha\beta}$.  The bulk susceptibility is $\chi = \chi_{cc} + 2\chi_{c\!f} + \chi_{f\!f}$. For $T>T^*$, $\chi_{c\!f}$  and $\chi_{cc}$ can be neglected, therefore $K_i = K_i^0 + B_i\chi$.  In this case $K_i$ is also linearly proportional to $K_j$: $K_i = a + b K_j$, where
\begin{eqnarray}
a &=& K_i^0 - (B_i/B_j) K_j^0\\
b &=& B_i/B_j.
\label{eqn:constants}
\end{eqnarray}
These relationships enables us to extract $\chi_{c\!f}$ using just  two pairs of Knight shifts  without the need for independent measurements of $\chi$.
Using Eqs. \ref{eqn:Knightshift} and \ref{eqn:constants} we find:
\begin{eqnarray}
\nonumber K_{c\!f}(T) &=& K_i(T) - a-b K_j(T) \\
&=& \left(A_i - \frac{B_i}{B_j}A_j\right)(\chi_{c\!f}(T) + \chi_{cc}(T)).
\end{eqnarray}
Since the hyperfine couplings are temperature independent, and $\chi_{cc}$ can be neglected, this quantity is proportional to $\chi_{c\!f}$ and becomes non-zero below $T^*$.


\bibliography{CeIrIn5_angulardependKnightshiftV7.bbl}

\end{document}